\documentclass[twocolumn,showpacs,amsfonts,aps,prc,nofootinbib,floatfix]{revtex4}

\usepackage{bm}
\usepackage{graphicx}
\usepackage{amsmath}

\def\la{\langle}
\def\ra{\rangle}
\def\dla{\la\!\la}
\def\dra{\ra\!\ra}

\def\eq{{\,=\,}}

\newcommand{\re}{\mbox{Re}}
\newcommand{\im}{\mbox{Im}}

\newcommand{\twopart}{\frac{dN}{d^3p_1d^3p_2}}
\newcommand{\onepart}[1]{\frac{dN}{d^3 #1}}
\newcommand{\avg}[1]{\left< #1 \right>} % for average
\newcommand{\sh}{\mbox{sh}}
\newcommand{\ch}{\mbox{ch}}

\begin{document}

%%%%%%%%%%%%%%%%%%%%%%%%Front Matter%%%%%%%%%%%%%%%%%%%%%%%%%%%%%%%%%%
%%%%%%%%%%%%%%%%%%%%%%%%%%%%%%%%%%%%%%%%%%%%%%%%%%%%%%%%%%%%%%%%%%%%%%

\title{HBT interferometry relative to the triangular flow plane in heavy-ion collisions}

\author{Christopher J. Plumberg}
\author{Chun Shen}
\author{Ulrich Heinz}
\affiliation{Department of Physics, The Ohio State University,
  Columbus, OH 43210-1117, USA}

\begin{abstract}
The PHENIX Collaboration has reported third-order harmonic oscillations of the source radius parameters when measuring the Hanbury Brown-Twiss correlation function for charged hadrons relative to the triangular flow angle. We explore possible origins of such third-order oscillations with a simple Gaussian source featuring both a triangular geometric deformation and triangular flow.  Third-order oscillations of the HBT radii can arise from a purely geometric triangular deformation superimposed on an azimuthally symmetric radial flow, or from a radially symmetry spatial distribution which expands anisotropically with a triangular component in the flow velocity profile.  In both cases the final particle momentum distribution features triangular flow.  We show that the two alternatives can be distinguished experimentally through the phase of the azimuthal oscillations of the HBT radii relative to the triangular flow plane.
\end{abstract}


\date{\today}

\maketitle

%%%%%%%%%%%%%%%%%%%%%%%%%%%%%%%%%%%%%%%%%%%%%%%%
\section{Introduction}
\label{sec1}
%%%%%%%%%%%%%%%%%%%%%%%%%%%%%%%%%%%%%%%%%%%%%%%%
\vspace*{-2mm}

Hanbury-Brown--Twiss (HBT) interferometry \cite{Heinz:1999rw} (also known as {\em femtoscopy} \cite{Lisa:2005dd,Lisa:2008gf,Lisa:2011na}) has become an indispensable tool in the description and understanding of heavy-ion collisions.  The observed particle output from heavy-ion collisions consists primarily of hadrons, whose momentum spectra and correlations contain information about the size and shape of the emission region when the particles last interacted with each other.  This last scattering is commonly referred to as (kinetic) freeze-out, and the space-time locations of the last scatterings define a (generically fuzzy) ``surface of last scattering'' or ``freeze-out surface".  Combined analyses of the observed particle momentum spectra and correlations yield not only geometric but also dynamical information about the fireball at freeze-out \cite{Heinz:1999rw}.  Because HBT interferometry yields both dynamical and geometric information, it provides a crucial window into the structure and evolution of the collision and its aftermath.

Azimuthally sensitive HBT (asHBT) analyses, where one studies the variation of the HBT correlator as a function of the emission direction perpendicular to the beam axis \cite{Voloshin:1995mc,Wiedemann:1997cr,Lisa:2000ip,Heinz:2002sq,Heinz:2002au,Retiere:2003kf}, are especially useful for probing anisotropies in the structure of the freeze-out surface.  It was shown in \cite{Lisa:2000ip,Heinz:2002sq} that first and second order harmonic oscillations of the HBT radii are dominated by geometric deformations of the emitting source relative to the beam direction and reaction plane.  Traditionally, one performs the asHBT analysis as a function of the pair emission angle $\Phi$ relative to the {\em elliptic flow angle} (or second-order {\em event plane angle} \cite{Lisa:2000xj,Adams:2003ra}) $\Psi_2$.  Dynamical models for the evolution of the fireball created in the collision allow to relate these observed geometric deformations in the final state to the initial conditions of the fireball.  For such a program of constraining from final state spectra and femtoscopic correlations the initial state of the fireball and its subsequent evolution, a proper understanding of the asHBT formalism is of paramount importance.

Recently the PHENIX collaboration \cite{TakafumiNiidaforthePHENIX:2013lia} performed the first experimental analysis of the azimuthal dependence of the HBT radii relative to the {\em triangular} flow plane $\Psi_3$. (For an earlier theoretical study see \cite{Voloshin:2011mg}.)
Triangular flow is dominated by event-by-event fluctuations in the initial transverse density profile \cite{Alver:2010gr} and largely independent of the collision centrality \cite{ALICE:2011ab,Chatrchyan:2012wg,ATLAS:2012at}.  In contrast to elliptic flow, which is strongly correlated with the direction of the impact parameter \cite{Qin:2010pf,Qiu:2011iv}, the direction $\Psi_3$ of triangular flow (the ``triangular flow plane'') is randomly distributed relative to the reaction plane \cite{Qin:2010pf,Qiu:2011iv,ALICE:2011ab}.  In two-pion correlations from central ($0{-}10\%$ centrality) Au+Au collisions at $\sqrt{s_\mathrm{NN}}=200$\,GeV, PHENIX \cite{TakafumiNiidaforthePHENIX:2013lia} saw clear third-order harmonic oscillations of the HBT radii relative to the triangular flow plane. The origin of these oscillations has not yet been understood.  In this paper, we perform a model study to qualitatively explore several possible explanations of the observations in \cite{TakafumiNiidaforthePHENIX:2013lia}. We show how the data distinguish between these alternatives, thereby favoring one specific scenario. A more quantitative analysis, based on a realistic hydrodynamic model for the expanding fireball created in these Au+Au collsions, will be published separately.

The outline of this paper is as follows.  In the next section, we generalize the theoretical formalism of asHBT to higher order harmonic oscillations.  In Section~\ref{sec3} we illustrate the general formalism for a simple Gaussian toy model and use it to obtain several qualitative results that appear to be robust and should survive a later analysis of realistic hydrodynamic sources.  Our results and conclusions about the meaning of the experimental data are summarized in Sec.~\ref{sec4}. 

%%%%%%%%%%%%%%%%%%%%%%%%%%%%%%%%%%%%%%%%%%%%%%%%%%
\vspace*{-2mm}
\section{Azimuthally sensitive femtoscopy}%: formalism}
\label{sec2}
\vspace*{-2mm}
%%%%%%%%%%%%%%%%%%%%%%%%%%%%%%%%%%%%%%%%%%%%%%%%%%
\subsection{Basics}
\label{sec2a}
\vspace*{-2mm}
%%%%%%%%%%%%%%%%%%%%%%%%%%%%%%%%%%%%%%%%%%%%%%%%%%
%
We study the two-particle correlation function
\begin{equation}
  C(\bm{p}_1, \bm{p}_2) \equiv \frac{E_1 E_2\twopart}{\left(E_1 \onepart{p_1}\right)\left(E_2 \onepart{p_2}\right)}.
\label{correlation_function_definition}
\end{equation}  
Its dependence on the momenta $\bm{p}_1,\,\bm{p}_2$ of the two measured particles can be expressed through the pair momentum $\bm{K}\eq(\bm{p}_1{+}\bm{p}_2)/2$ and their relative momentum $\bm{q}\eq\bm{p}_1{-}\bm{p}_2$, $ C(\bm{q}, \bm{K})$. Assuming a ``chaotic source'' that emits the two particles independently (i.e. without dynamical correlations) and ignoring final state interactions between the two particles after their last scattering with the source medium, the two-particle phase-space distribution for the emitted particles factorizes, and for a pair of identical bosons this correlation function can be written as \cite{Heinz:1999rw,Lisa:2005dd,Chapman:1994ax}
\begin{eqnarray}
\label{eq2}
  C(\bm{q},\bm{K}) &=& 1 + \frac{\left|\int d^4x\, S(x,K)\, e^{i q \cdot x}\right|^2}
                                              {\int d^4x\, S(x,K{+}q/2) \int d^4y\, S(y,K{-}q/2)}
\nonumber\\
  &\approx& 1 + \left | \frac{\int d^4x\, S(x,K)\, e^{i q \cdot x}}
                                         {\int d^4x\, S(x,K)} \right |^2. 
\end{eqnarray}  
Here the emission function $S(x,K)$ is the one-particle Wigner density (i.e. the quantum mechanical phase space distribution) of the emitted particles, whose space-time integral gives the single-particle momentum distribution:
\begin{equation}
\label{eq3}
  E_{p} \onepart{p} = \int d^4x\, S(x,p).
\end{equation}
The last step in Eq.~(\ref{eq2}) uses the ``smoothness approximation'' $S(x,K{\pm}q/2){\,\approx\,}S(x,K)$ in the $q$-range where the correlation function deviates from 1 (which can be justified \cite{Pratt:1997pw} for sufficiently large sources). $q$ and $K$ are four-vectors, with energy components $q^0\eq{E}_1{-}E_2\eq\bm{\beta}\cdot\bm{q}$ (where $\bm{\beta}\eq\bm{K}/K^0$) and $K^0\eq(E_1{+}E_2)/2$ (which for massive particles with Compton wave length smaller than the source size can be taken as approximately on-shell, $K^0{\,\approx\,}E_K\eq\sqrt{\bm{K}^2{+}m^2}$, such that $\bm{\beta}$ becomes the velocity of the pair).

For Gaussian sources whose spatial emission region, for each value $\bm{K}$ of the pair momentum, can be completely characterized by its spatial variances, the correlator $C(\bm{q},\bm{K})$ has a Gaussian $\bm{q}$-dependence \cite{Chapman:1994ax}:
\begin{equation}
\label{eq4}
  C(\bm{q},\bm{K}) = 1 + \exp\Bigl[-\!\!\!\!\sum_{i,j=o,s,l} q_i q_j R^2_{ij}(\bm{K})\Bigr]. 
\end{equation} 
Here
\begin{equation}
\label{eq5}
  R^2_{ij}(\bm{K}) = \avg{(\tilde{x}_i-\beta_i \tilde{t})(\tilde{x}_j-\beta_j \tilde{t})},\quad i,j=o,s,l, 
\end{equation}  
where $o,s,l$ denote the outward (pointing along the pair momentum $\bm{K}_\perp$ in the transverse plane), sideward (pointing perpendicular to $\bm{K}_\perp$ in the transverse plane), and longitudinal direction along the beam. $\avg{\dots}$ stands for the average over the emission function:
\begin{equation}
\label{eq6}
   \avg{f(x)} \equiv \frac{\int d^4x\, f(x)\, S(x,K)}{\int d^4x \, S(x,K)}.
\end{equation} 
The parameters $R^2_{ij}(\bm{K})$ are called ``HBT radii'' and represent the widths of the effective emission region (``homogeneity region'' \cite{Makhlin:1987gm}) for particles with momentum $\bm{K}$. $\bar{x}^\mu(\bm{K}){\,\equiv\,}\avg{x^\mu}(\bm{K})$ is the center of that homogeneity region (for a smooth source, this is close to the point of highest emissivity for particles with momentum $\bm{K}$), and $\tilde{x}_{\mu}{\,\equiv\,}x^\mu{-}\avg{x^\mu}(\bm{K})$ in Eq.~(\ref{eq5}) denotes the distance from that point. Note that the off-diagonal components $R^2_{ij}$ ($i{\neq}j$) need not be positive. In this paper, we focus on the $\bm{K}$ dependence of just two of these radius parameters, $R^2_s(\bm{K})$ and $R^2_o(\bm{K})$.

%%%%%%%%%%%%%%%%%%%%%%%%%%%%%%%%%%%%%%%%%%%%%%%%%%
\subsection{Femtoscopy of fluctuating sources}
\label{sec2b}
%%%%%%%%%%%%%%%%%%%%%%%%%%%%%%%%%%%%%%%%%%%%%%%%%%

Extracting all source information contained in $C(\bm{q},\bm{K})$ requires to fully explore its 6-dimensional momentum dependence. Due to the limited number of final state particles, this is not possible for a single heavy-ion collision. Experimental HBT analyses are therefore based on a statistical average over large numbers (typically millions) of collision events. We now know that, even for tightly defined collision centrality, the initial density profile of the collision fireball, and therefore its final-state emission function $S(x,K)$, fluctuates from collision to collision. Its regions of homogeneity and thus its HBT radii are therefore stochastically fluctuating quantities. In the past, theoretical analyses of HBT data have implicitly assumed that the entire ensemble of collision events underlying the measurement can be characterized by a single ``average emission function''
\begin{equation}
\label{eq7}
   \bar{S} (x, K) \equiv \langle S(x, K) \rangle_\mathrm{ev},
\end{equation}
where $\langle\dots\rangle_\mathrm{ev}$ stands for the average over an ensemble of collision events. More correctly, however, the measured correlator is obtained from ensemble-averaged true and mixed-event pair distributions:
\begin{equation}
\label{eq8}
  \avg{C(\bm{p}_1, \bm{p}_2)}_\mathrm{ev} = 
  \frac{\left.\avg{\twopart\right|_\mathrm{true}}_\mathrm{ev}}
         {\left.\avg{\twopart\right|_\mathrm{mixed}}_\mathrm{ev}}.
\end{equation}  
The numerator consists of pairs with both particles taken from the same event, the pairs in the denominator mix particles from different events. The denominator factorizes into a product of ensemble-averaged single particle spectra,
\begin{equation}
\label{eq9}
  \avg{C(\bm{p}_1, \bm{p}_2)}_\mathrm{ev} = 
  \frac{\avg{\twopart}_\mathrm{ev}}
         {\avg{\onepart{p_1}}_\mathrm{ev}\avg{\onepart{p_2}}_\mathrm{ev}},                                                 
\end{equation}  
corresponding to the theoretical expression
\begin{eqnarray}
\label{eq10}
 && \langle C(\bm{q}, \bm{K}) \rangle_\mathrm{ev} = 
  1 + \frac{\avg{\vert\int d^4x\, e^{iq \cdot x} S(x, K) \vert^2}_\mathrm{ev}}  
               {\vert\int d^4x\, \langle  S(x, K) \rangle_\mathrm{ev} \vert^2}
\nonumber\\
   \neq&& \bar{C}(\bm{q}, \bm{K}) \equiv 
   1 + \frac{\vert\int d^4x\, e^{iq \cdot x} \bar{S}(x, K) \vert^2}{\vert\int d^4x\, \bar{S}(x, K) \vert^2}.                                                                          
\end{eqnarray}
So the correlator $\langle C(\bm{q}, \bm{K}) \rangle_\mathrm{ev}$ measured in the event ensemble does not agree with the correlator $\bar{C}(\bm{q}, \bm{K})$ corresponding to the ensemble-averaged emission function. Writing the emission function $S(x,K)$ of a given single event as the ensemble-averaged $\bar{S}$ plus a fluctuation $\delta S$,
\begin{equation}
\label{eq11}
  S(x, K) = \bar{S}(x, K) + \delta S(x, K) 
  \mbox{, with } \langle \delta S(x, K) \rangle_\mathrm{ev} = 0,
\end{equation}
the numerator in the first line of Eq.~(\ref{eq10}) separates into
\begin{eqnarray}
\label{eq12}
  \left\langle \left\vert \int d^4x\, e^{iq\cdot x} S(x, K) \right\vert^2 \right\rangle_\mathrm{ev} 
  \!\!\!\!\!
  &&= \left\vert \int d^4x\, e^{i q \cdot x} \bar{S}(x, K) \right\vert^2 
\\ \notag
  +&& \left\langle \left\vert \int d^4 x e^{i q \cdot x} \delta S(x, K) 
          \right\vert^2 \right\rangle_\mathrm{ev}
\end{eqnarray}
such that
\begin{eqnarray}
\label{eq13}
  \avg{C(\bm{q}, \bm{K})}_\mathrm{ev} = \bar{C}(\bm{q}, \bm{K}) + 
  \frac{\avg{\vert \int d^4x\, e^{i q \cdot x} \delta S(x, K)\vert^2}_\mathrm{ev}}
         {\vert \int d^4 x \bar{S}(x, K) \vert^2}.
  \nonumber\\
\end{eqnarray}
We denote the second term by $I(\bm{q},\bm{K})$; it gives the contribution to the measured two-particle correlation function arising from event-by-event fluctuations of the emission function. Its $q\to0$ limit
\begin{eqnarray}
\label{eq14}
  I(\bm{q}{=}0,\bm{K}) &=& 
  \frac{\avg{\vert \int d^4x\, \delta S(x, K) \vert^2}_\mathrm{ev}}
         {\vert \int d^4 x \bar{S}(x, K) \vert^2} 
\notag \\
&=& \frac{\avg{\left(\frac{d\delta N}{dY K_\perp dK_\perp d\Phi}\right)^2}_\mathrm{ev}} 
               {\avg{\frac{dN}{dY K_\perp dK_\perp d\Phi}}_\mathrm{ev}^2}
\end{eqnarray}
is the normalized variance of the single particle spectrum at momentum $\bm{K}{\,\equiv\,}(K_\perp, \Phi, Y)$. This term is positive, causing $\avg{C(\bm{q}, \bm{K})}_\mathrm{ev}$ to approach an intercept slightly larger than 2 as $\bm{q}$ goes to zero.\footnote{We here 
   consider the idealized case, ignoring long-lived resonance decays and other effects that 
   tend to reduce $\avg{C(0,\bm{K})}_\mathrm{ev}$ \cite{Heinz:1999rw,Lisa:2005dd}.} 
For large $q$ the numerator of the last term in Eq.~(\ref{eq13}) oscillates to zero. A more detailed investigation of the effect of $I(\bm{q},\bm{K})$ on the HBT radii and the $\bm{q}=0$ intercept of the correlator are left for a separate study. We here focus on the properties of the ensemble-averaged emission function $\bar{S}(x,K)$ and its correlator $\bar{C}(\bm{q},\bm{K})$, dropping the bars over $\bar{S}$ and $\bar{C}$ for simplicity.

%%%%%%%%%%%%%%%%%%%%%%%%%%%%%%%%%%%%%%%%%%%%%%%%%%
\subsection{$\bm{n}^\mathrm{th}$-order flow angles}
\label{sec2c}
%%%%%%%%%%%%%%%%%%%%%%%%%%%%%%%%%%%%%%%%%%%%%%%%%%

We are interested in the dependence of the correlator $C(\bm{q},\bm{K})$ and its associated HBT radii on the azimuthal angle $\Phi$ in which the particle pair is emitted. Due to the random orientation of the collision fireball in the transverse plane, only the relative angle between $\Phi$ and some direction associated with the event orientation has physical meaning. Experimentally, the orientation of each event can be characterized by its harmonic flow angles $\Psi_n$. They are defined through the expectation value of $e^{in\Phi}$ with the single-particle spectrum $E_K\onepart{K}$ \cite{Heinz:2013th}:
\begin{eqnarray}
\label{eq15}
  v_n e^{i n \Psi_n} \equiv 
  \frac{\int^{\infty}_0 dK_T K_T\int_{-\pi}^{\pi}d\Phi\, e^{i n \Phi}\int d^4 x \, S(x,K)}{\int^{\infty}_0 dK_T K_T\int_{-\pi}^{\pi}d\Phi\,\int d^4 x \, S(x,K)}. 
\end{eqnarray}
$v_n$ is the associated $n^\mathrm{th}$-order harmonic flow coefficient. $\Psi_2$ and $\Psi_3$ define the elliptic and triangular flow planes, respectively, for each event. In this work we study the azimuthal dependence of the HBT radii as a function of $\Phi{-}\Psi_n$, in particular for $n\eq2$ and 3.

%%%%%%%%%%%%%%%%%%%%%%%%%%%%%%%%%%%%%%%%%%%%%%%%%%
\subsection{Fourier expansion of the emission function and HBT radii}
\label{sec2d}
%%%%%%%%%%%%%%%%%%%%%%%%%%%%%%%%%%%%%%%%%%%%%%%%%%

To simplify notation, we introduce the normalized emission function
\begin{equation}
\label{eq16}
   \tilde{S}(x,K) \equiv \frac{S(x,K)}{\int d^4x\,S(x,K)}, \quad \int d^4x\,\tilde{S}(x,K)=1.
\end{equation}
We use polar Milne coordinates, $x^\mu\eq(\tau,r,\phi,\eta)$ and $K^\mu\eq(M_\perp,K_\perp,\Phi,Y)$, where $x{\eq}r\cos\phi$, $y{\eq}r\sin\phi$, $\tau=\sqrt{t^2{-}z^2}$, $\eta\eq\frac{1}{2}\ln\left(\frac{t{+}z}{t{-}z}\right)$, $K_x{=}{K}_\perp\cos\Phi$, $K_y{=}{K}_\perp\sin\Phi$, $M_\perp\eq\sqrt{K_\perp^2{+}m^2}$, and $Y\eq\frac{1}{2}\ln\left(\frac{E_K{+}K_z}{E_K{-}K_z}\right)$.

We perform a double Fourier expansion in both the spatial angle $\phi$ associated with $\bm{r}$ and the momentum angle $\Phi$ associated with $\bm{K}_\perp$:
\begin{eqnarray}
\label{eq17}
  \!\!\!\!\!&&\tilde{S}(x,K) = \!\!\!\!\!\! \sum^{\infty}_{l,m=-\infty} \!\!\!\!\!\!
  S_{l,m}(r,\tau,\eta;K_\perp,Y)\, e^{-il\phi - im(\Phi{-}\Psi_n)},
\nonumber \\ 
  \!\!\!\!\!&&S_{l,m} = e^{-i m \Psi_n}\int^{\pi}_{-\pi} \frac{d\phi}{2\pi} e^{il\phi}  
                                              \int^{\pi}_{-\pi} \frac{d\Phi}{2\pi} e^{im\Phi} \tilde{S}(\phi,\Phi) .
\end{eqnarray}
Note that the first (second) index labels the spatial (momentum) Fourier component. Since $S(x,K)$ is real, the Fourier coefficients satisfy
\begin{eqnarray}
\label{eq18}\!\!\!\!\!\!\!
  &&S_{l,m}=S^*_{-l,-m},
\\\nonumber
 &&S_{l,m}{+}S_{-l,-m}\eq2\,\re\, S_{l,m},\ 
%\nonumber\\
 S_{l,m}{-}S_{-l,-m}\eq2\,i\,\im\, S_{l,m}.
\end{eqnarray}
The Fourier expansion in $\Phi{-}\Psi_n$ permits decomposition of the HBT radii into oscillations of different harmonic order:
\begin{eqnarray}
\label{eq19}
  R^2_{ij}(\bm{K)} &=&  R^2_{ij,0}  + 2 \sum^{\infty}_{k=1} 
  \Bigl( R^{2(c)}_{ij,k}\cos[k(\Phi{-}\Psi_n)]
\nonumber \\
   &&\qquad\qquad\quad
      +\, R^{2(s)}_{ij,k}\sin[k(\Phi{-}\Psi_n)] \Bigr)\qquad
\end{eqnarray}
where the coefficients $R^{2(c,s)}_{ij,k}$ are functions of $K_\perp$ and $Y$.

To derive mathematical expressions for the oscillation amplitudes $R^{2(c,s)}_{ij,k}$ we recall that $R_{ij}^2\eq\avg{(\tilde x_i{-}\beta_i \tilde t)^2}$ can be written explicitly as follows:
\begin{eqnarray}
\label{eq20}
  &&R_s^2 = \avg{x_s^2}-\avg{x_s}^2,
\nonumber\\
  &&R_o^2 = \avg{(x_o{-}\beta_\perp\tau\ch\eta)^2}-\avg{x_o{-}\beta_\perp \tau\ch\eta}^2,
\nonumber\\
  &&R_\ell^2 = \avg{\tau^2(\sh\eta{-}\beta_\ell\ch\eta)^2}
                     - \avg{\tau(\sh\eta{-}\beta_\ell\ch\eta)}^2,
\nonumber\\
  &&R_{os}^2 = \avg{(x_o{-}\beta_\perp\tau\ch\eta)x_s} - \avg{x_o{-}\beta_\perp\tau\ch\eta}\avg{x_s},
\nonumber\\ \!\!\!\!
  &&R_{\ell s}^2 = \avg{\tau(\sh\eta{-}\beta_\perp\ch\eta)x_s} - \avg{\tau(\sh\eta{-}\beta_\perp\ch\eta)}\avg{x_s},\quad
\nonumber\\
  &&R_{o\ell}^2 = \avg{(x_o{-}\beta_\perp\tau\ch\eta)\tau(\sh\eta{-}\beta_\perp\ch\eta)} 
\nonumber\\
  &&\qquad\  \ - \avg{x_o{-}\beta_\perp\tau\ch\eta} \avg{\tau(\sh\eta{-}\beta_\perp\ch\eta)}.
\end{eqnarray}
Here
\begin{equation}
\label{eq21}
   x_o = r \cos(\phi{-}\Phi), \quad x_s = r \sin(\phi{-}\Phi).
\end{equation}
We write
\begin{equation}
  \tilde S = \sum_{l{=}{-}\infty}^\infty {\mathcal{Z}}_l\,e^{-il(\phi{-}\Phi)},
\end{equation}
where
\begin{equation}
\label{eq22}
  {\mathcal{Z}}_l \equiv e^{-il\Psi_n} \!\!\!\!
  \sum_{m=-\infty}^{\infty} \!\! S_{l,m-l}\, e^{-im(\Phi{-}\Psi_n)}
  \equiv {\mathcal{X}}_l + i {\mathcal{Y}}_l.
\end{equation}
Clearly, ${{\mathcal{Z}}_l}^*\eq{\mathcal{Z}}_{-l}$, and ${\mathcal{X}}_l,\, {\mathcal{Y}}_l$ are real functions of $(r,\tau,\eta;K_\perp,Y)$ satisfying ${\mathcal{X}}_l\eq{\mathcal{X}}_{-l}$, ${\mathcal{Y}}_l\eq{-}{\mathcal{Y}}_{-l}$. Introducing the shorthand notation
\begin{equation}
\label{eq23}
  \int_x \equiv \int_{-\infty}^\infty d\eta \int_0^\infty \tau\,d\tau \int_0^\infty r\, dr,
\end{equation}
we find
\begin{eqnarray}
\label{eq24}
  &&\avg{x_s} = \int_x\! 2\pi r {\mathcal{Y}}_1, \quad 
       \avg{x_o} = \int_x\! 2\pi r {\mathcal{X}}_1,
\nonumber\\ 
  &&\avg{z} = \int_x\! 2\pi \tau\sh\eta {\mathcal{X}}_0, \quad 
       \avg{t} = \int_x\! 2\pi \tau\ch\eta {\mathcal{X}}_0,
\end{eqnarray}
and
\begin{eqnarray}
\label{eq25}
  &&\avg{x^2_s} = \int_x\! \pi r^2 ({\mathcal{X}}_0{-}{\mathcal{X}}_2),\ 
      \avg{x^2_o} = \int_x\! \pi r^2 ({\mathcal{X}}_0{+}{\mathcal{X}}_2),
\nonumber\\
  &&\avg{z^2} = \int_x\! 2\pi \tau^2 \sh^2\eta \,{\mathcal{X}}_0,\ 
       \avg{t^2} = \int_x\! 2\pi \tau^2 \ch^2\eta \,{\mathcal{X}}_0,
\nonumber\\
  &&\avg{x_o x_s} = \int_x\! \pi r^2 \,{\mathcal{Y}}_2,\ 
       \avg{x_o t} = \int_x\! 2\pi r \tau \ch\eta \,{\mathcal{X}}_1,
\\
  &&\avg{x_s t} = \int_x\! 2\pi r \tau \ch\eta \,{\mathcal{Y}}_1,
      \avg{x_o z} = \int_x\! 2\pi r \tau \sh\eta \,{\mathcal{X}}_1,
\nonumber\\ \nonumber\!\!\!\!\!\!
  &&\avg{x_s z} = \int_x\! 2\pi r \tau \sh\eta \,{\mathcal{Y}}_1,
       \avg{z t} = \int_x\! 2\pi \tau^2 \sh\eta\ch\eta \,{\mathcal{X}}_0.\ 
\end{eqnarray}
We now write
\begin{equation}
\label{eq26}
  {\mathcal{Z}}_l \equiv 
  \sum_{m=-\infty}^{\infty} z_l^m\, e^{-im(\Phi{-}\Psi_n)},
\end{equation}
remembering that $l$ is the spatial Fourier index and $m$ labels the harmonic in the emission angle $\Phi$. We note that
\begin{equation}
\label{eq27}
  z_l^m = x_l^m+iy_l^m\equiv e^{-il\Psi_n} \, S_{l,m-l} = \bigl(z_{-l}^{-m}\bigr)^*,
\end{equation}
which leads to the decomposition
\begin{eqnarray}
\label{eq28}
   {\mathcal{X}}_l \!\!&=&\!\! x_l^0 + 2\!\sum_{k=1}^\infty\! 
   \left[ x_l^k\cos\bigl(k(\Phi{-}\Psi_n)\bigr) +
           y_l^k\sin\bigl(k(\Phi{-}\Psi_n)\bigr) \right],
\nonumber\\
   {\mathcal{Y}}_l \!\!&=&\!\! y_l^0 + 2\!\sum_{k=1}^\infty\! 
   \left[ y_l^k\cos\bigl(k(\Phi{-}\Psi_n)\bigr) -
           x_l^k\sin\bigl(k(\Phi{-}\Psi_n)\bigr) \right].
\nonumber\\                                                    
\end{eqnarray}                                                   
Equation~(\ref{eq28}) immediately yields the contributions to the oscillation amplitudes $R^{2(c,s)}_{ij,k}$ in Eq.~(\ref{eq19}) that arise from the expressions listed in (\ref{eq25}). The contributions from the terms involving products of the mean coordinates listed in Eq.~(\ref{eq24}) require a little more effort. We define
\begin{eqnarray}
\label{eq29}
   \xi^k_{ll'}&=&\frac{1}{4}\sum_{m=-\infty}^\infty \bigl(z_l^m{+}z_{-l}^m\bigr)
                                                                       \bigl({z'}_{\!\!l'}^{k-m}{+}{z'}_{\!\!-l'}^{k-m}\bigr),
\nonumber\\
   \zeta^k_{ll'}&=&\frac{1}{4i}\sum_{m=-\infty}^\infty \bigl(z_l^m{+}z_{-l}^m\bigr)
                                                                       \bigl({z'}_{\!\!l'}^{k-m}{-}{z'}_{\!\!-l'}^{k-m}\bigr),
\nonumber\\
   \tau^k_{ll'}&=&-\frac{1}{4}\sum_{m=-\infty}^\infty \bigl(z_l^m{-}z_{-l}^m\bigr)
                                                                       \bigl({z'}_{\!\!l'}^{k-m}{-}{z'}_{\!\!-l'}^{k-m}\bigr),
\end{eqnarray}
where the prime on $z'$ indicates that it depends on primed spatial coordinates $(r',\tau',\eta')$.
These quantities satisfy $(\xi^k_{ll'})^*\eq\xi^{-k}_{ll'}$, $(\zeta^k_{ll'})^*\eq\zeta^{-k}_{ll'}$, and $(\tau^k_{ll'})^*\eq\tau^{-k}_{ll'}$. After some algebra we find
\begin{eqnarray}
\label{eq30}
  {\mathcal{X}}_l{\mathcal{X}}'_{l'} &=&  \xi^0_{ll'} + 2 \sum^{\infty}_{k=1} 
  \left[\re\xi^k_{ll'}{\cdot}\cos\bigl(k(\Phi{-}\Psi_n)\bigr)\right.
\nonumber \\
   &&\qquad\qquad\ 
      \left. +\ \im\xi^k_{ll'}{\cdot}\sin\bigl(k(\Phi{-}\Psi_n)\bigr) \right],
\nonumber\\
  {\mathcal{X}}_l{\mathcal{Y}}'_{l'} &=&  \zeta^0_{ll'} + 2 \sum^{\infty}_{k=1} 
  \left[\re\zeta^k_{ll'}{\cdot}\cos\bigl(k(\Phi{-}\Psi_n)\bigr)\right.
\nonumber \\
   &&\qquad\qquad\ 
      \left. +\ \im\zeta^k_{ll'}{\cdot}\sin\bigl(k(\Phi{-}\Psi_n)\bigr)\right],
\nonumber\\
  {\mathcal{Y}}_l{\mathcal{Y}}'_{l'} &=&  \tau^0_{ll'} + 2 \sum^{\infty}_{k=1} 
  \left[\re\tau^k_{ll'}{\cdot}\cos\bigl(k(\Phi{-}\Psi_n)\bigr)\right.
\nonumber \\
   &&\qquad\qquad\ 
     \left. +\ \im\tau^k_{ll'}{\cdot}\sin\bigl(k(\Phi{-}\Psi_n)\bigr)\right].\qquad
\end{eqnarray}
Inserting Eqs.~(\ref{eq28},\ref{eq30}) into Eqs.~(\ref{eq24},\ref{eq25}) and further into Eqs.~(\ref{eq20}) yields the desired Fourier expansions (\ref{eq19}) for the HBT radii. For the sideward and outward radius parameters $R_s^2$ and $R_o^2$ these read explicitly as follows (explicit expressions for the remaining HBT radius parameters can be found in Appendix~\ref{appa}):
\begin{eqnarray}
\label{eq31}
  R_s^2 &=& \int_x \pi r^2({\mathcal{X}}_0{-}{\mathcal{X}}_2) 
  - \int_{x,x'} 4\pi^2 r r' {\mathcal{Y}}_1{\mathcal{Y}}'_1,
\end{eqnarray}
with (see Eq.~(\ref{eq19}))
\begin{eqnarray}
\nonumber
  R_{s,0}^2 &=& \int_x \pi r^2 (x_0^0{-}x_2^0) - \int_{x,x'} 4\pi^2 r r' \tau_{11}^0,
\\ \nonumber
  R_{s,k}^{2(c)} &=& \int_x \pi r^2 (x_0^k{-}x_2^k) - \int_{x,x'} 4\pi^2 r r' \re\tau_{11}^k,
\end{eqnarray}
\begin{equation}
\label{eq32}
  R_{s,k}^{2(s)} = \int_x \pi r^2 (y_0^k{-}y_2^k) - \int_{x,x'} 4\pi^2 r r' \im\tau_{11}^k,
\end{equation}
and
\begin{widetext}
\begin{eqnarray}
\label{eq33}
  R_o^2 &=& \int_x \left[\pi r^2({\mathcal{X}}_0{+}{\mathcal{X}}_2)
                             -4\pi r \beta_\perp \tau\ch\eta {\mathcal{X}}_1
                             +2\pi\beta_\perp^2\tau^2\ch^2\eta{\mathcal{X}}_0\right]
\nonumber\\
  && -\int_{x,x'} \left[4\pi^2 r r' {\mathcal{X}}_1{\mathcal{X}}'_1 
                              -8\pi^2 r \beta_\perp\tau'\ch\eta'{\mathcal{X}}_1{\mathcal{X}}'_0
                          +4\pi^2 \beta_\perp^2\tau\tau'\ch\eta\,\ch\eta'{\mathcal{X}}_0{\mathcal{X}}'_0
                          \right],
\\
\label{eq34}
  R_{o,0}^2 &=& \int_x \left[\pi r^2 (x_0^0{+}x_2^0) - 4\pi r\beta_\perp\tau\ch\eta\, x_1^0
                                         +2\pi\beta_\perp^2\tau^2\ch^2\eta x_0^0\right]
\nonumber\\
  && - \int_{x,x'} \left[ 4\pi^2 r r' \xi_{11}^0 -8\pi^2 r\beta_\perp\tau'\ch\eta'\xi_{10}^0
                                +4\pi^2\beta_\perp^2\tau\tau'\ch\eta\,\ch\eta'\xi_{00}^0\right],
\nonumber\\
  R_{o,k}^{2(c)} &=& \int_x \left[\pi r^2 (x_0^k{+}x_2^k) - 4\pi r\beta_\perp\tau\ch\eta\, x_1^k
                                         +2\pi\beta_\perp^2\tau^2\ch^2\eta x_0^k\right]
\nonumber\\
  && 
  - \int_{x,x'} \left[ 4\pi^2 r r' \re\xi_{11}^k -8\pi^2 r\beta_\perp\tau'\ch\eta'\re\xi_{10}^k
                                +4\pi^2\beta_\perp^2\tau\tau'\ch\eta\,\ch\eta'\re\xi_{00}^k\right],
\nonumber\\
  R_{o,k}^{2(s)} &=& \int_x \left[\pi r^2 (y_0^k{+}y_2^k) - 4\pi r\beta_\perp\tau\ch\eta\, y_1^k
                                         +2\pi\beta_\perp^2\tau^2\ch^2\eta\, y_0^k\right]
\nonumber\\
  && - \int_{x,x'} \left[ 4\pi^2 r r' \im\xi_{11}^k -8\pi^2 r\beta_\perp\tau'\ch\eta'\im\xi_{10}^k
                                +4\pi^2\beta_\perp^2\tau\tau'\ch\eta\,\ch\eta'\im\xi_{00}^k\right].
\end{eqnarray}
\end{widetext}

Already here a first important conclusion can be drawn. Remember that in Eq.~(\ref{eq22}) the subscript $l$ on ${\mathcal{Z}}_l$ is the spatial Fourier index on the involved Fourier components $S_{l,m{-}l}$. Eqs.~(\ref{eq20},\ref{eq24}) show that the HBT radii can be written entirely in terms of harmonic source coefficients with spatial Fourier indices $l\in(0,1,2)$; 
conspicuously absent are any spatial harmonics with $l{\,\geq\,}3$. This is ultimately a consequence of the definition of the HBT radii in terms of linear and quadratic space-time moments of the source. They can directly probe geometric fireball deformations only up to second harmonic order. The oscillations in $R^2_s$ and $R^2_o$ are therefore insensitive to spatial harmonics of order $l{\,\geq\,}3$ in the source function. In particular, for a stationary (non-expanding) fireball, any spatial deformations of higher than second harmonic order (e.g. any triangular geometric deformation of the source) do not contribute to the oscillation amplitudes $R^{2(c,s)}_{ij,k}$ \cite{Voloshin:2011mg}. Only if the spatial deformation couples to a collective flow pattern (which can be radially symmetric), thereby producing a momentum anisotropy of harmonic order $m{\,\geq\,}3$, will we see azimuthal oscillations of the HBT radii with higher than second harmonic frequency.

This implies that the measurement of a triangular oscillation of the HBT radii cannot be automatically identified with the spatial triangularity at freeze-out. This is different from elliptic HBT oscillations: there it was shown that, for particles pairs with small pair momentum $K_\perp{\,\to\,}0$, geometric deformation effects dominate over anisotropic flow effects \cite{Lisa:2000ip,Heinz:2002sq}, and the following relation between the oscillation amplitude of the square of the sideward radius $R_s^2$ and final source ellipticity holds: $\epsilon_\mathrm{final} \approx 2R^2_{s,2}/R^2_{s,0}$ \cite{Retiere:2003kf,Adams:2003ra}. Eqs.~(\ref{eq20},\ref{eq24}) imply that a similar relation cannot hold for the triangular freeze-out eccentricity.

We also see from the equations above that the oscillation amplitudes $R^{2(c,s)}_{ij,k}$ receive contributions from all harmonic orders $m$ in the momentum angle $\Phi-\Psi_n$. Therefore there is no restriction on which flow harmonics (i.e. which types of anisotropies in the collective flow pattern of the source) can contribute to oscillations of the HBT radii. 

Summarizing these observations we conclude that a triangular oscillation in the $R^2_{ij}$ may originate from at least three possible sources: (1) a non-zero triangular flow anisotropy in a spatially isotropic source; (2) a non-zero triangular eccentricity coupling to azimuthally symmetric radial flow; or (3) a combination of the two. In addition, higher-order flow harmonics can couple non-linearly to other spatial deformation components to produce third-order oscillations in $\Phi$. In the next section we explore possibilities (1)-(3) with a toy model.

%%%%%%%%%%%%%%%%%%%%%%%%%%%%%%%%%%%%%%%%%%%%%%%%%%
\section{Triangular oscillations from a toy model}
\label{sec3}
%%%%%%%%%%%%%%%%%%%%%%%%%%%%%%%%%%%%%%%%%%%%%%%%%%

Motivated by the PHENIX data in \cite{TakafumiNiidaforthePHENIX:2013lia}, we study in this section azimuthal oscillations of the sideward and outward HBT radii, $R_s^2$ and $R_o^2$, relative to the triangular flow angle $\Psi_3$. As already mentioned, we will focus on the correlator $\bar{C}(\bm{q}, \bm{K})$ associated with the ensemble-averaged emission function
$\bar{S}(x,K)$. Since triangular flow is driven by event-by-event fluctuations \cite{Qin:2010pf,Qiu:2011iv}, to get a nonzero triangular spatial deformation of $\bar{S}(x,K)$ we must orient all events such that their triangular flow planes align before taking the average. As a result of this rotation, the ellipticities of the individual events will end up randomly oriented (since the elliptic and triangular event planes are uncorrelated \cite{Alver:2010gr,ALICE:2011ab,Qin:2010pf,Qiu:2011iv,Qiu:2012uy,Jia:2012sa}), and the ensemble-averaged source is expected to exhibit no net elliptic deformation, at any impact parameter. We will therefore model $\bar{S}(x,K)$ to feature triangular deformations in both space and hydrodynamic flow velocity, but no other anisotropies. Motivated by hydrodynamics, we will assume that any triangular hydrodynamic flow points in the direction of steepest descent of the density profile, i.e. the third-order participant plane angle $\Phi_3$ (pointing in the direction of one of the tips of the triangular spatial density distribution) and the hydrodynamic flow angle $\bar{\psi}_3$ (pointing in the direction of largest hydrodynamic flow) are out of phase by $\pi/3$ \cite{Qin:2010pf,Qiu:2011iv}. This allows for some relaxation of any initial triangular spatial source deformation before kinetic freeze-out, in response to triangular hydrodynamic flow, but not to the extent that the source ``overshoots'' the spatially azimuthally symmetric state and freezes out with a triangularly deformed final density distribution whose tips point {\em into} the direction of the triangular flow.  

%%%%%%%%%%%%%%%%%%%%%%%%%%%%%%%%%%%%%%%%%%%%%%%%%%
\subsection{The model emission function}
\label{sec3a}
%%%%%%%%%%%%%%%%%%%%%%%%%%%%%%%%%%%%%%%%%%%%%%%%%%

In this spirit, we generalize the well-studied hydrodynamically expanding Gaussian source model used in Refs.~\cite{Csorgo:1994fg,Chapman:1994ax,Wiedemann:1997cr,Heinz:2002au,Retiere:2003kf}) as follows:
\begin{eqnarray}
\label{eq35}
&&\tilde{S}(x,K) = \frac{S_0(K)}{\tau}
           \exp\left[ -\frac{(\tau{-}\tau_f)^2}{2(\Delta\tau)^2} 
                          -\frac{(\eta{-}\eta_0)^2}{2(\Delta\eta)^2} \right. 
\nonumber \\ \nonumber
	& & \qquad\qquad\qquad\qquad\qquad
	-\frac{r^2}{2R^2}\left( 1+2\bar{\epsilon}_3\cos(3(\phi{-}\bar{\psi}_3)) \right)  
\\ 
        &&\ \  \left. - \frac{1}{T_0}\Bigl(M_{\perp}\ch(\eta{-}Y)\ch \eta_t 
	              - K_{\perp} \cos(\phi{-}\Phi)\sh\eta_t\Bigr) \right]\!,
\end{eqnarray} 
where $S_0(K)$ normalizes $\tilde{S}(x,K)$ to $\int d^4x\,\tilde{S}(x,K)\eq1$. 

The first two lines specify a freeze-out density distribution that is Gaussian in longitudinal proper time and space-time rapidity and an azimuthally modulated Gaussian in the transverse coordinate $r$. The third line specifies the momentum structure of the source in terms of a flow-boosted Boltzmann factor $\exp[-p{\cdot}u(x)/T_0]$ where the flow four-velocity $u^\mu(x)$ is parametrized in terms of a $(\tau,\eta)$-independent transverse flow rapidity profile:
\begin{equation}
\label{eq36}
  \eta_t(r,\phi) = \eta_f\frac{r}{R}\left( 1+2\bar{v}_3 \cos(3(\phi{-}\bar{\psi}_3)) \right).
\end{equation}  
This source depends on the following parameters: $\tau_f$ is the mean emission (or freeze-out)time for the observed particles, $\Delta\tau$ the emission duration. We assume $\Delta\tau{\,\ll\,}\tau_f$ such that time integrals $\int_0^\infty d\tau\dots$ can be replaced by $\int_{-\infty}^\infty d\tau\dots$ without large error. In the calculations below we use $\tau_f\eq10$\,fm/$c$ and $\Delta\tau\eq1$\,fm/$c$. $\eta_0$ is the mean space-time rapidity of the emitted particles;
for Au+Au collisions measured in the CM frame, $\eta_0\eq0$ is the appropriate choice. $\Delta\eta$ describes the width of the freeze-out space-time rapidity distribution; we assume approximately boost-invariant longitudinal flow such that, on average, $\eta$ equals the final momentum rapidity of the measured particles. For pions from 200\,$A$\,GeV Au+Au collisions, the width of the measured rapidity distribution has been determined as $\sigma_\pi{\,\approx\,}2.3$ \cite{Bearden:2004yx}, so we take $\Delta\eta\eq2.3$. We compute 2-pion correlations at midrapidity, so we set $m\eq{m}_\pi\eq139$\,MeV and $Y\eq0$.

$R$ is the mean two-dimensional Gaussian source radius at freeze-out; for central Au+Au collisions, we take $R\eq4.5$\,fm. The kinetic freeze-out temperature is set to $T_0\eq120$\,MeV. $\bar{\epsilon}_3$ parametrizes the triangular spatial source deformation which will be varied. The transverse flow rapidity $\eta_t$ is taken to grow linearly with $r$, with average strength $\eta_f\eq0.6$ at $r\eq{R}$, and modulated azimuthally by a triangular flow whose strength, parametrized by $\bar{v}_3$, will be varied. The transverse flow rapidity is largest in direction $\phi\eq\bar\psi_3$ which we call the hydrodynamic triangular flow direction. The second line in Eq.~(\ref{eq35}) shows that the effective source radius is smallest in direction $\bar\psi_3$, so $\bar\psi_3$ also denotes the direction of steepest descent of the spatial density profile.

%
%=================== Fig. 1 ====================
\begin{figure}[htb]
    \includegraphics[width=\linewidth]{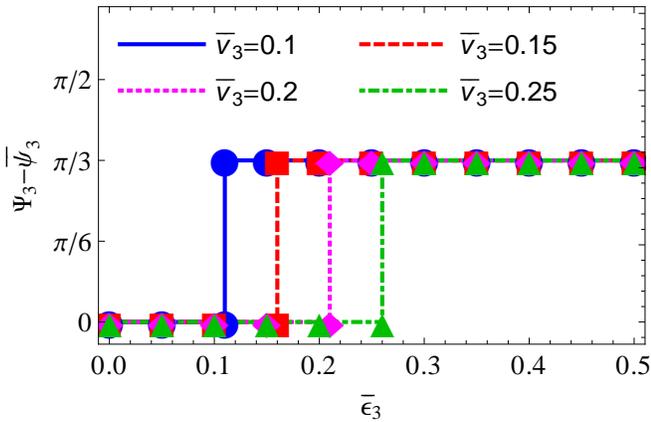}
    \caption{(Color online)
    The difference $\Psi_3{-}\bar{\psi}_3$ between the triangular flow angle $\Psi_3$ of the 
    emitted particles and the direction $\bar{\psi}_3$ of the triangular hydrodynamic flow 
    anisotropy, as a function of the geometric triangularity $\bar\epsilon_3$, for several values 
    of the magnitude of the triangular deformation of the flow velocity, $\bar{v}_3$. The critical 
    value of $\bar\epsilon_3$ where $\Psi_3{-}\bar{\psi}_3$ flips by $\pi/3$ is positively correlated 
    with $\bar{v}_3$ but insensitive to the strength $\eta_f$ of the radial flow. 
    Due to the symmetry of our toy model source function (\ref{eq35}), 
    $\dla\sin\bigl(3(\Phi{-}\bar{\psi}_3)\bigr)\dra\eq0$ always (see Eq.~(\ref{eq15})),
    and the sign of $\dla\cos\bigl(3(\Phi{-}\bar{\psi}_3)\bigr)\dra$ distinguishes between
    flow angles $\Phi_3\eq\bar{\psi}_3$ and $\Phi_3\eq\bar{\psi}_3{\,\pm\,}\frac{\pi}{3}$. 
    \label{F1}
    }
\end{figure}
%=============================================
%

It is important to conceptually differentiate between $\bar\psi_3$, the direction of the hydrodynamic triangular flow in the source, and $\Psi_3$, the direction of the triangular flow angle extracted via Eq.~(\ref{eq15}) from the final particle momentum spectrum. To illustrate the issue, consider a source with non-zero radial flow $\eta_f$ modulated by a small anisotropy $\bar{v}_3$ that we hold fixed, and a variable spatial eccentricity $\bar\epsilon_3$. For $\bar\epsilon_3\eq0$ (i.e. an azimuthally symmetric spatial density distribution), the non-zero $\bar{v}_3$ obviously induces triangular flow in direction $\bar\psi_3$, hence $\Psi_3(\bar\epsilon_3{=}0)\eq\bar\psi_3$. As we increase $\bar\epsilon_3$, the effective spatial source radius is reduced in direction $\bar\psi_3$ and increased along the directions $\bar\psi_3{\,\pm\,}\frac{\pi}{3}$. Due to the linear radial flow profile, this reduces the transverse flow in direction $\bar\psi_3$ while enhancing it in the directions $\bar\psi_3{\,\pm\,}\frac{\pi}{3}$. Eventually this stronger flow in directions $\bar\psi_3{\,\pm\,}\frac{\pi}{3}$ overwhelms the triangular flow in direction $\bar\psi_3$, and $\Psi_3(\bar\epsilon_3\,\mathrm{large})\eq\bar\psi_3{\,\pm\,}\frac{\pi}{3}$. This flip of $\Psi_3$ by $\pi/3$ as a function of geometric source triangularity is illustrated in Fig.~\ref{F1}. As we will see, it is the root source of our ability to distinguish experimentally between triangular HBT oscillations driven by triangular geometric deformations vs. those driven by triangular anisotropies in the hydrodynamic flow.

%%%%%%%%%%%%%%%%%%%%%%%%%%%%%%%%%%%%%%%%%%%%%%%%%%
\subsection{HBT oscillations from the toy model}
\label{sec3b}
%%%%%%%%%%%%%%%%%%%%%%%%%%%%%%%%%%%%%%%%%%%%%%%%%%

\begin{figure}[b]
    \includegraphics[width=\linewidth]{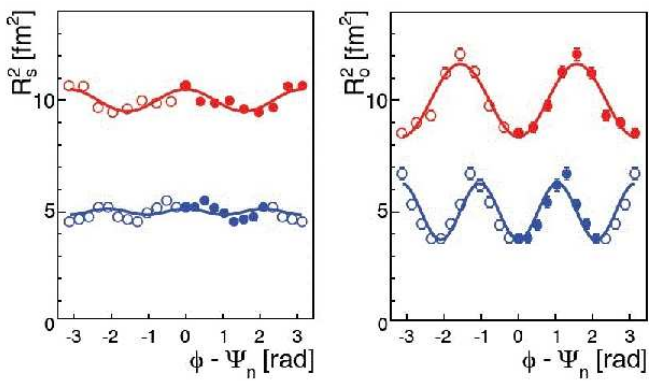}
    \caption{(Color online) Second and third order oscillations of $R^2_s$ and $R^2_o$ 
    measured by the PHENIX Collaboration in central ($0{-}10\%$) 200\,$A$\,GeV Au+Au 
    collisions \cite{TakafumiNiidaforthePHENIX:2013lia}. For better visibility, the average 
    values $R^2_{s,0}$, $R^2_{o,0}$ of the two radius parameters were set by hand 
    to 5 and 10\,fm$^2$, respectively, when plotting the third- and second-order oscillations.
    %The actually measured values are $R^2_{s,0}{\,\simeq\,}19$\,fm$^2$ and 
    %$R^2_{o,0}{\,\simeq\,}20.5$\,fm$^2$ \cite{TakafumiNiidaforthePHENIX:2013lia}, with
    %$\sim8\%$ systematic errors. [Chris, I don't know where you found that info -- it's not
    %in the cited reference.]
    \label{F2}
    }
  \end{figure}

The toy model study presented in this paper was motivated by recent experimental data from the PHENIX Collaboration, shown by T. Niida at the {\em Quark Matter 2012} conference \cite{TakafumiNiidaforthePHENIX:2013lia} and reproduced in Fig.~\ref{F2}. The data show clear 
%
%=================== Fig. 2 ====================
%=============================================
%
triangular oscillations as a function of the pair emission angle $\Phi$, with $R_s$ being maximal and $R_o$ minimal in triangular flow direction $\Psi_3$. For the selected almost central Au+Au collisions ($0{-}10\%$ centrality), the oscillation amplitude for $R_o^2$ is much larger than for $R_s^2$. 

As already discussed at the end of Sec.~\ref{sec2}, the observed small triangular oscillation amplitude $R_{s,3}^2$ of $R_s^2$ cannot \cite{Voloshin:2011mg} be directly interpreted as evidence for a small geometric triangularity of the source at freeze-out. What, then, is the correct interpretation of the experimental observations?

%
%=================== Fig. 3 ====================
\begin{figure}[ht]
    \includegraphics[width=\linewidth]{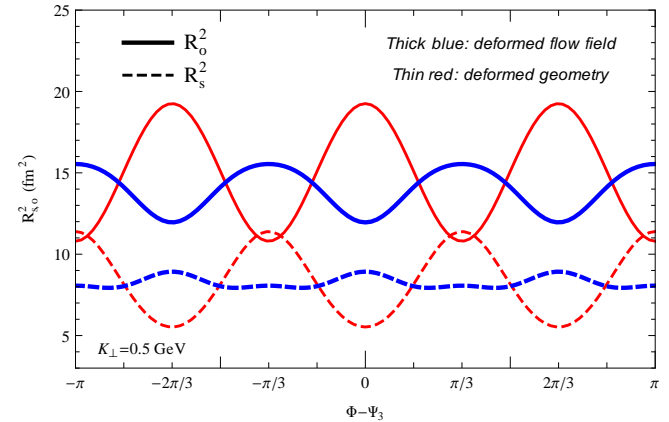}
    \caption{(Color online)
    Triangular oscillations of $R^2_s$ (dashed) and $R^2_o$ (solid) for pion pairs with 
    momentum $K_\perp\eq0.5$\,GeV, as a function of emission angle $\Phi$ relative to 
    the triangular flow direction $\Psi_3$. Shown are results for two model scenarios: A 
    deformed flow field ($\bar{v}_3\eq0.25$) in a spatially isotropic ($\bar\epsilon_3\eq0$)
    density distribution (thick blue lines), and a source with triangular geometric deformation
    ($\bar\epsilon_3\eq0.25$) expanding with radially symmetric ($\bar{v}_3\eq0$) flow (thin red
    lines). For the two scenarios the oscillations of both $R^2_s$ and $R^2_o$are seen to 
    be out of phase by $\pi/3$. 
    \label{F3}
    }
\end{figure}
%=============================================
%

In Fig.~\ref{F3} we show triangular oscillations relative to the triangular flow plane of $R_s^2$ and $R_o^2$ for pion pairs with $K_\perp\eq0.5$\,GeV from our toy model, for two opposite model assumptions: Thin red lines (solid for $R_o^2$, dashed for $R_s^2$) correspond to a triangular source with spatial deformation $\bar\epsilon_3\eq0.25$ expanding radially symmetrically ($\bar{v}_3\eq0$); in this case the triangular flow of the emitted hadrons is entirely due to the triangular geometric deformation which couples to the radial flow profile. Thick blue lines show the HBT radii from an azimuthally symmetric ($\bar\epsilon_3\eq0$) source density profile, superimposed by transverse flow with triangular anisotropy $\bar{v}_3\eq0.25$. We make several observations: (i) Due to the symmetry of the emission function, the coefficients $R_{s,3}^{2(s)}$ and $R_{o,3}^{2(s)}$ of the sine terms in Eq.~(\ref{eq19}) vanish; we therefore drop from hereon the superscript $(c)$ on the (non-vanishing) cosine amplitudes $R_{s,3}^{2(c)}$ and $R_{o,3}^{2(c)}$. (ii) For both sources, the oscillation amplitudes are larger in the outward than in the sideward direction. (iii) In both cases, the outward and sideward HBT radii oscillate out of phase by $\pi/3$. (i)--(iii) are in qualitative agreement with the experimental data. (iv) For both $R_s^2$ and $R_o^2$, the oscillations for the geometrically deformed source and for the source with only a deformed flow profile are out of phase by $\pi/3$. Only for the flow-anisotropy-dominated case (thick red lines) do the HBT radii oscillate in phase with the experimental observations.

\begin{figure}[htb]
    \includegraphics[width=\linewidth]{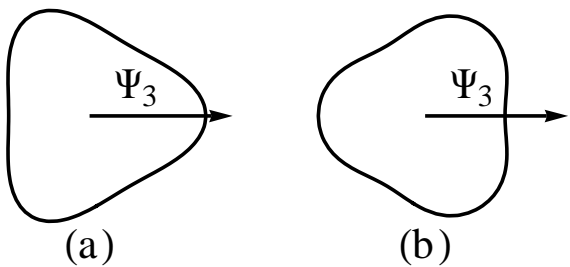}
    \caption{
    Half-maximum contour plots for the emission functions for $K_\perp\eq0$ pions, 
    for two sources with $\bar\epsilon\eq0.25$, $\bar{v}_3\eq0$ (``geometry dominated'', left)
    and with $\bar\epsilon\eq0$, $\bar{v}_3\eq0.25$ (``flow anisotropy dominated'', right). In
    both cases the sources are oriented such that the triangular flow angle $\Psi_3$ points 
    in $x$ direction.
    \label{F4}}
\end{figure}

Fig. ~\ref{F4} illustrates that, in a coordinate system whose $x$ axis points along the triangular flow vector such that $\Psi_3\eq0$, the sources for these two scenarios appear rotated by $60^\circ$ relative to each other. The left sketch shows the 50\% contour of the emission function for pions with $K_\perp\eq0$ for the source with $\bar\epsilon_3\eq0.25$, $\bar{v}_3\eq0$ while the source corresponding to the right sketch has $\bar\epsilon\eq0$ and $\bar{v}_3\eq0.25$.

%
%=================== Fig. 4 ====================
%=============================================
%

\begin{figure}[htb]
    \includegraphics[width=\linewidth]{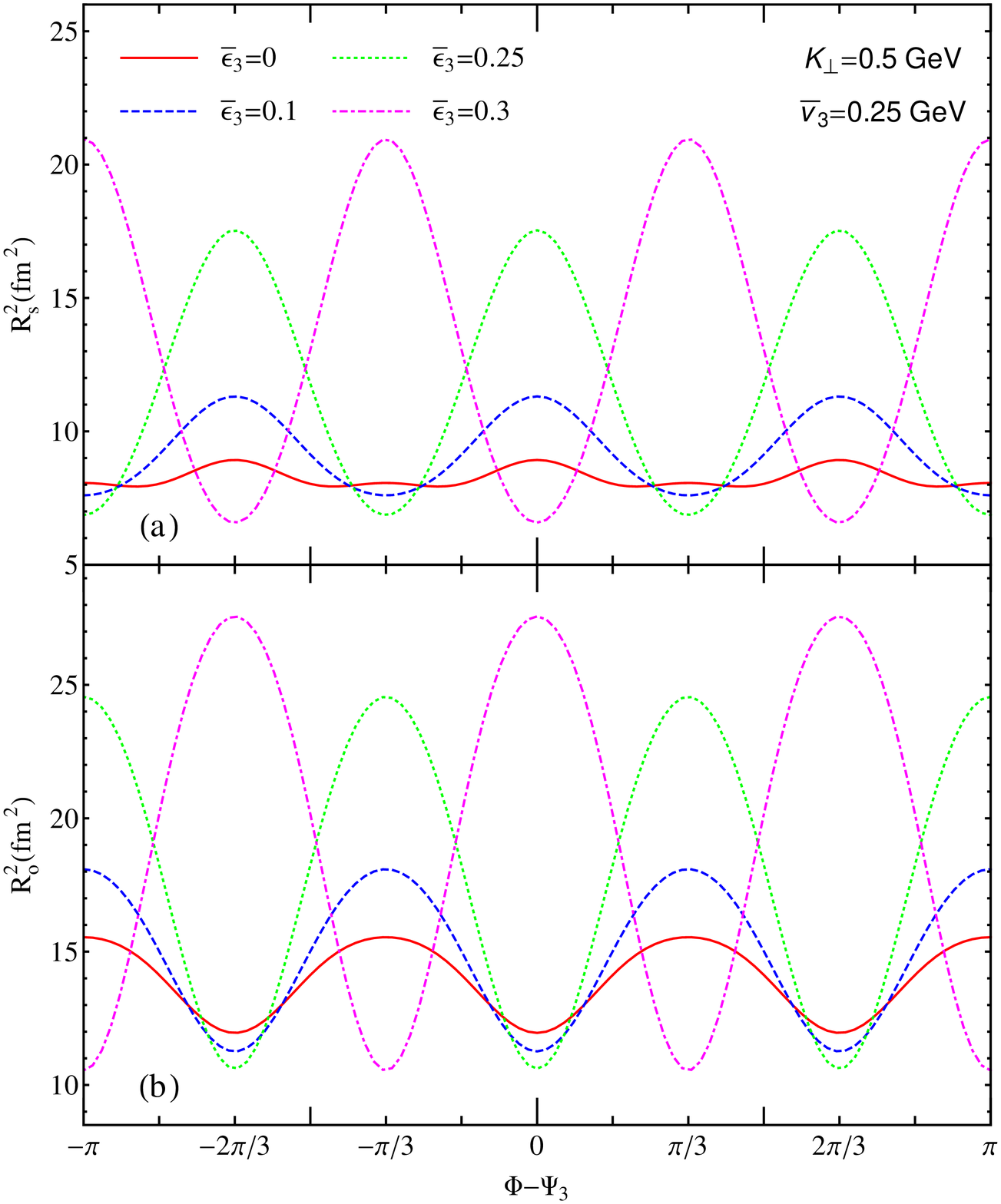}
    \caption{(Color online)
    Triangular oscillations of $R^2_s$ (a) and $R^2_o$ (b) for pion pairs with 
    momentum $K_\perp\eq0.5$\,GeV, as a function of emission angle $\Phi$ relative to 
    the triangular flow direction $\Psi_3$. Shown are results for a source with fixed triangular
    flow anisotropy $\bar{v}_3\eq0.25$, for a range of triangular spatial deformations 
    $\bar\epsilon_3$. One sees that the phase of the HBT oscillations relative to the flow angle 
    $\Psi_3$ flips by $\pi/3$ between $\bar\epsilon_3=0.25$ and 0.3, as a result $\Psi_3$ itself
    flipping by $\pi/3$ (see Fig.~\ref{F1}). 
    \label{F5}
    }
\end{figure}

Of course, in general the source will exhibit flow and geometric anisotropies concurrently. Figure~\ref{F5} shows the triangular oscillations of $R^2_{o,s}$ for a sequence of sources that %
%=================== Fig. 5 ====================
%=============================================
%
all have the same triangular flow anisotropy $\bar{v}_3\eq0.25$ but feature varying degrees of spatial triangularity $\bar\epsilon_3$. The oscillation amplitudes, as well as the mean values, of both $R_s^2$ and $R_o^2$ increase monotonically with triangularity $\bar\epsilon_3$. As Fig.~\ref{F1} shows, just after $\bar\epsilon_3$ reaches the value 0.26, the flow angle $\Psi_3$ flips from 0 to $\pi/3$. This is reflected in Fig.~\ref{F5} by a sudden $60^\circ$ phase shift relative to $\Psi_3$ of both $R_s^2$ and $R_o^2$ oscillations. For the given flow anisotropy $\bar{v}_3\eq0.25$, as long as $\bar\epsilon_3{\,<\,}0.26$, $R_o^2$ has a minimum for emission along the triangular flow plane, as observed in experiment; only for spatial deformation $\bar\epsilon_3{\,>\,}0.26$ the outward radius becomes maximal for emission in $\Psi_3$ direction. The observed phase of the sideward and outward HBT thus tells us only that the measured HBT oscillations correspond to a source in which triangular flow anisotropies dominate over geometric triangular deformation effects that are boosted by superimposed radial flow. Thus, while a direct measurement of $\bar\epsilon_3$ is not possible, the observed oscillation phase can perhaps be used to put limits on the ratio $\bar\epsilon_3/\bar{v}_3$ in theoretical models. It is, however, likely that such limits depend on the details of the model emission function.    

To explore this last question a bit further, we 
%
%=================== Fig. 6 ====================
\begin{figure}[b]
    \includegraphics[width=\linewidth]{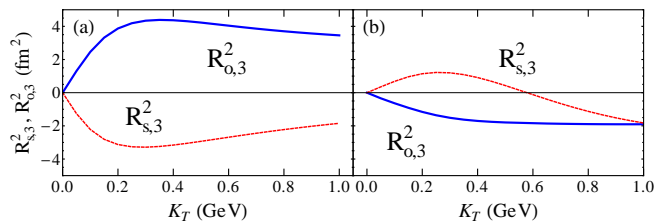}
    \caption{(Color online) 
    $K_\perp$-dependence of the third-order oscillation amplitudes of $R^2_s$ and $R^2_o$,
    for the ``geometry dominated'' (a) and ``flow anisotropy dominated'' (b) sources studied
    in Figures~\ref{F3} and \ref{F4}. 
    \label{F6}}
\end{figure}
%=============================================
%
return to the completely ``geometry dominated'' ($\bar\epsilon\eq0.25,\ \bar{v}_3\eq0$) and completely ``flow anisotropy dominated'' ($\bar\epsilon\eq0,\ \bar{v}_3\eq0.25$) sources studied in Figs.~\ref{F3} and \ref{F4}. In Fig.~\ref{F6} we show for these extreme models the $K_\perp$-dependence of their third-order oscillation amplitudes $R_{s,3}^2$ and $R_{o,3}^2$. At low $K_\perp$, the opposite signs of the oscillation amplitudes in Figs.~\ref{F6}a and \ref{F6}b reflect the phase shift of the oscillations by $\pi/3$ between the two types of sources. For the flow-dominated source in Fig.~\ref{F6}b, the sign of $R_{s,3}^2$ flips near $K_\perp{\,\simeq\,}0.6$\,GeV, causing the squares of the sideward and outward radii to oscillate in phase at large $K_\perp$. This contradicts the out-of-phase oscillation pattern seen in the data (Fig.~\ref{F2}) which are integrated over $K_\perp$ and thus, due to the exponentially falling single-particle spectra, dominated by $K_\perp{\,\lesssim\,}0.5$\,GeV. On the other hand, the zero of $R_{s,3}^2$ in Fig.~\ref{F6}b near $K_\perp{\,\approx\,}0.6$\,GeV provides an intriguing possible explanation for the experimentally observed much smaller oscillation amplitude in sideward than in outward direction: $R_{s,3}^2$ passes through zero close to the most likely $K_\perp$ value in the measured pion sample. Obviously, it is unsafe to assume that the experimentally measured source is characterized by zero triangularity $\bar\epsilon_3$ as assumed in Fig.~\ref{F6}b. However, this figure strongly suggests that a careful experimental study of the $K_\perp$-dependence of the HBT oscillation amplitudes can provide significant quantitative constraints on the relative importance of spatial triangularity and triangular flow anisotropy in heavy-ion collision fireballs at freeze-out.

%%%%%%%%%%%%%%%%%%%%%%%%%%%%%%%%%%%%%%%%%%%%%%%%%
\subsection{Emission contours as functions of emission angle}
\label{sec3c}
%%%%%%%%%%%%%%%%%%%%%%%%%%%%%%%%%%%%%%%%%%%%%%%%%

%
%=================== Fig. 7 ====================
\begin{figure}[htb]
    \includegraphics[width=\linewidth]{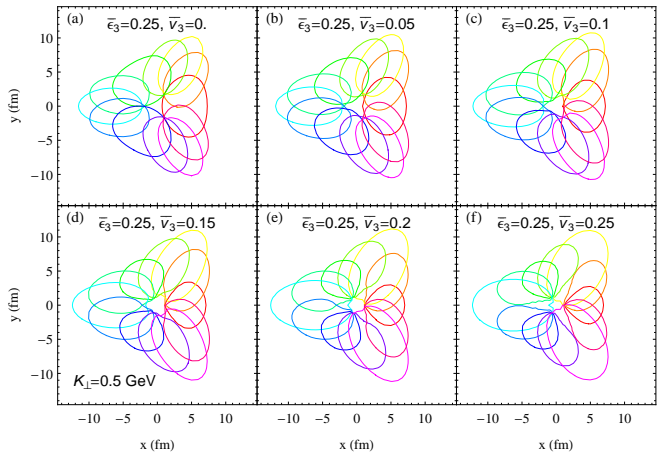}
    \caption{(Color online) 
    Contours of the homogeneity regions for pion pairs with $K_\perp\eq0.5$\,GeV,
    as a function of the emission angle $\Phi{-}\Psi_3$ (increasing in steps of $30^\circ$),
    for ``geometry dominated'' sources with fixed $\bar\epsilon_3\eq0.25$ and varying
    $\bar{v}_3{\,\leq\,}\bar\epsilon_3$, as indicated in the panels.
    \label{F7}}
\end{figure}
%=============================================
%

We close this section by illustrating the characteristics of our toy model source, as seen by the HBT femtoscope, for a range of spatial triangularities $\bar\epsilon_3$ and triangular flow anisotropies $\bar{v}_3$, by plotting the 50\% contours of the effective emission regions (``regions of homogeneity'') as functions of emission angle $\Phi{-}\Psi_3$. We always orient the sources such that $\Psi_3\eq0$, and focus on pion pairs with $K_\perp\eq0.5$\,GeV.

%
%=================== Fig. 8 ====================
\begin{figure}[htb]
    \includegraphics[width=\linewidth]{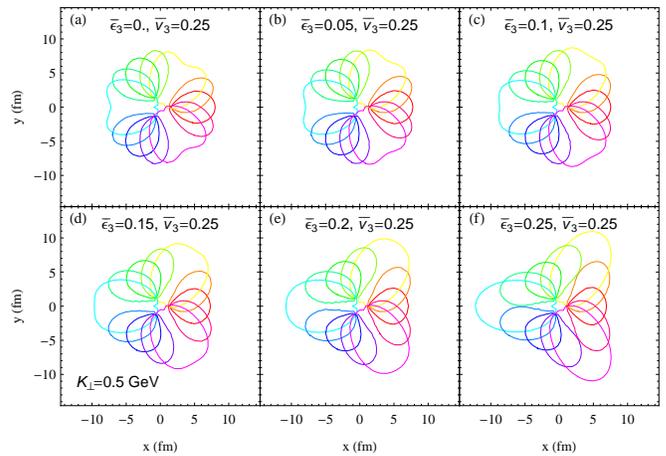}
    \caption{(Color online) 
    Similar to Fig.~\ref{F7}, but for ``flow anisotropy dominated'' sources with fixed 
    $\bar{v}_3\eq0.25$ and varying $\bar{\epsilon}_3{\,\leq\,}\bar{v}_3$, as indicated 
    in the panels.
    \label{F8}}
\end{figure}
%=============================================
%
%
%=================== Fig. 9 ====================
\begin{figure}[htb]
    \includegraphics[width=\linewidth]{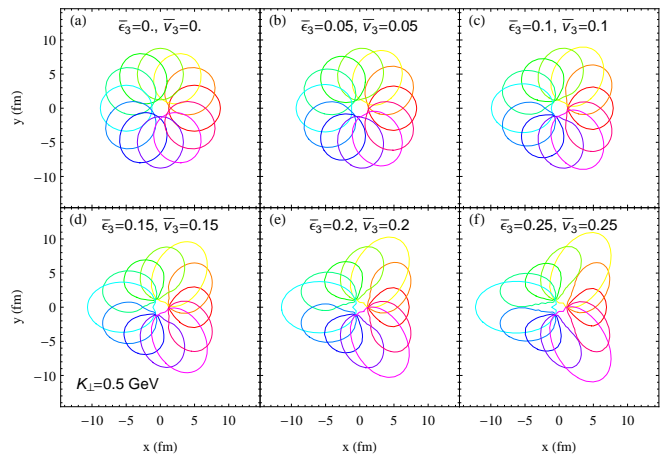}
    \caption{(Color online) 
    Similar to Figs.~\ref{F7} and \ref{F8}, but for sources with varying but equal values of
    $\bar{\epsilon}_3$ and $\bar{v}_3$, as indicated in the panels.
    \label{F9}}
\end{figure}
%=============================================
%

In Fig.~\ref{F7} we study sources with fixed spatial triangularity $\bar\epsilon_3\eq0.25$ and variable triangular flow anisotropy $0{\,\leq\,}\bar{v}_3{\,\leq\,}0.25$. In Fig.~\ref{F8} the roles of $\bar\epsilon_3$ and $\bar{v}_3$ are reversed: $\bar{v}_3$ is fixed at 0.25, and $\bar\epsilon_3$ is varied between 0 and 0.25. In Fig.~\ref{F9}, finally, we set $\bar\epsilon_3\eq\bar{v}_3$ and vary both parameters together from 0 to 0.25. The emission angle $\Phi$ is increased in steps of $\pi/6\eq30^\circ$.

By following the emission contours around the circle, focussing on the $\Phi$-dependence of their diameters in radial and tangential directions, one can qualitatively see the oscillations of the HBT radii in outward and sideward directions. Obviously, the flow-dominated sources with $\bar{v}_3{\,>\,}\bar\epsilon_3$ generate the most complicated shapes for the homogeneity regions. For the sources with $\bar{v}_3\eq\bar\epsilon_3$ in Fig.~\ref{F9}, the homogeneity regions are nearly elliptical for almost all parameter values and emission angles.

%%%%%%%%%%%%%%%%%%%%%%%%%%%%%%%%%%%%%%%%%%%%%%%%%
\section{Conclusions}
\label{sec4}
%%%%%%%%%%%%%%%%%%%%%%%%%%%%%%%%%%%%%%%%%%%%%%%%%

Using a simple triangularly deformed Gaussian toy model for an expanding source with triangular flow anisotropy, we performed an exploratory study of the possible origins
of the third-order harmonic oscillations of the HBT radius parameters as functions of the azimuthal emission angle around the beam axis that were recently discovered by the PHENIX collaboration in 200\,$A$\,GeV Au+Au collisions at RHIC. We developed a general double Fourier expansion technique to study HBT oscillations of any harmonic order and applied the technique to our toy model for the case of triangular ($n\eq3$) oscillations. We also pointed out that, since the oscillations are measured relative to the triangular flow which is generated by fluctuations in the initial colliding nucleon distribution around the smooth average Woods-Saxon density, one should in principle account for event-by-event fluctuations of the emitting source, and that the HBT radii of the ensemble-averaged emission function are not identical with those extracted from experiment by averaging the correlation function over pairs from a large ensemble of events. Exploring this theme further will be left to a future study. Here we focused on harmonic oscillations of the HBT radii from a smooth (but deformed) average source.

We showed that, in contrast to second-order oscillations which, in certain limits, can measure the elliptic geometric deformation of the source, third-order oscillations in general provide no direct information about the triangular geometric shape of the emitting source. A spatial triangular deformation can cause third-order HBT oscillations only by coupling to radial transverse flow, and this complicates their interpretation because the HBT radii depend also on the amount of radial flow. Perhaps more naturally, third-order HBT oscillations can be driven by triangular hydrodynamic flow. We showed that these two alternatives generate triangular flows of the emitted particles whose directions differ by $60^\circ$, and that this causes a similar $60^\circ$ phase shift of the HBT radii oscillations when measured as functions of the emission angle relative to the triangular flow direction. The phase of the oscillations seen in the PHENIX experiment was shown to be consistent with a source whose flow anisotropy $\bar{v}_3$ dominates over its spatial triangularity $\bar\epsilon_3$ in its effect on azimuthal HBT variations. Furthermore, we showed that in such a ``flow anisotropy dominated'' scenario the oscillation amplitude for the square of the sideward HBT radius $R_s^2$ flips sign at intermediate $K_\perp$ values of order 0.5\,GeV, providing a natural explanation for the observed weakness of the sideward oscillations relative to those in outward direction. A systematic study of the $K_\perp$-dependence of the HBT oscillation amplitudes can help to substantiate this interpretation of the data and constrain the relative magnitude of the triangular spatial deformation $\bar\epsilon_3$ and flow anisotropy $\bar{v}_3$.
 
\bigskip

%%%%%%%%%%%%% Acknowledgements %%%%%%%%%%%%%%%%%%%%%%%%%%%
\acknowledgements
\vspace*{-2mm}
We thank Zhi Qiu for many fruitful discussions. This work was supported by the U.S.\ Department of Energy under Grants No.~\rm{DE-SC0004286} and (within the framework of the JET Collaboration) \rm{DE-SC0004104}.

\appendix

%%%%%%%%%%%%% Appendix A %%%%%%%%%%%%%%%%%%%%%%%%%%%
\section{Fourier expansion of additional $R^2_{ij}$} \label{appa}
%%%%%%%%%%%%%%%%%%%%%%%%%%%%%%%%%%%%%%%%%%%%%%%

Here we complement the Fourier expansions for $R^2_s$ and $R^2_o$ given in Eqs.~(\ref{eq31}-\ref{eq34}) by those for the remaining four HBT radius parameters:

\begin{widetext}
\begin{eqnarray}
\label{eqA1}
  R_{\ell}^2 &=& \int_x 2 \pi \tau^2 \left(\sh\eta{-}\beta_{\ell} \ch\eta \right)^2{\mathcal{X}}_0
   -\int_{x,x'} 4\pi^2 \tau \tau' \left( \sh\eta{-}\beta_{\ell} \ch\eta \right)
                          \left( \sh\eta'{-}\beta_{\ell} \ch\eta' \right) {\mathcal{X}}_0{\mathcal{X}}'_0,
\\
\label{eqA2}
  R_{\ell,0}^2 &=& \int_x 2 \pi \tau^2 \left(\sh\eta{-}\beta_{\ell} \ch\eta \right)^2 x^0_0
   -\int_{x,x'} 4\pi^2 \tau \tau' \left( \sh\eta{-}\beta_{\ell} \ch\eta \right)
                          \left( \sh\eta'{-}\beta_{\ell} \ch\eta' \right) \xi_{00}^0,
\nonumber\\
  R_{\ell,k}^{2(c)} &=& \int_x 2 \pi \tau^2 \left(\sh\eta{-}\beta_{\ell} \ch\eta \right)^2 x^k_0
  -\int_{x,x'} 4\pi^2 \tau \tau' \left( \sh\eta{-}\beta_{\ell} \ch\eta \right)
                          \left( \sh\eta'{-}\beta_{\ell} \ch\eta' \right) \re\xi_{00}^k,
\nonumber\\
  R_{\ell,k}^{2(s)} &=& \int_x 2 \pi \tau^2 \left(\sh\eta{-}\beta_{\ell} \ch\eta \right)^2 y^k_0
   -\int_{x,x'} 4\pi^2 \tau \tau' \left( \sh\eta{-}\beta_{\ell} \ch\eta \right)
                          \left( \sh\eta'{-}\beta_{\ell} \ch\eta' \right) \im\xi_{00}^k
\end{eqnarray}
\begin{eqnarray}
\label{eqA3}
  R_{os}^2 &=& \int_x 2 \pi \left[r^2 {\mathcal{Y}}_2 - 2\beta_{\perp} \tau \ch\eta {\mathcal{Y}}_1
  							\right]
   -\int_{x,x'} 4\pi^2 \left[r r' {\mathcal{X}}_1{\mathcal{Y}}'_1 -\beta_{\perp} r' \tau \ch \eta
   						 {\mathcal{X}}_0{\mathcal{Y}}'_1
   						\right],
\\ \label{eqA4}
  R_{os,0}^2 &=& \int_x 2 \pi \left[r^2 y_2^0 - 2\beta_{\perp} \tau \ch\eta y_1^0
  							\right]
   -\int_{x,x'} 4\pi^2 r' \left[r \zeta_{11}^0 -\beta_{\perp} \tau \ch \eta \zeta_{01}^0
   						\right],
\nonumber\\
  R_{os,k}^{2(c)} &=& \int_x 2 \pi \left[r^2 y_2^k - 2\beta_{\perp} \tau \ch\eta y_1^k
  							\right]
  -\int_{x,x'} 4\pi^2 r' \left[r \re\zeta_{11}^k -\beta_{\perp} \tau \ch \eta \re\zeta_{01}^k
   						\right],
\nonumber\\
  R_{os,k}^{2(s)} &=& -\int_x 2 \pi \left[r^2 x_2^k - 2\beta_{\perp} \tau \ch\eta x_1^k
  							\right]
  -\int_{x,x'} 4\pi^2 r' \left[r \im\zeta_{11}^k -\beta_{\perp} \tau \ch \eta \im\zeta_{01}^k
   						\right]
\end{eqnarray}
\begin{eqnarray}
\label{eqA5}
  R_{\ell s}^2 &=& \int_x 2 \pi r \tau \left(\sh\eta{-}\beta_{\ell} \ch\eta \right){\mathcal{Y}}_1
   -\int_{x,x'} 4\pi^2 r' \tau \left( \sh\eta{-}\beta_{\ell} \ch\eta \right)
                          {\mathcal{X}}_0{\mathcal{Y}}'_1,
\\
\label{eqA6}
  R_{\ell s,0}^2 &=& \int_x 2 \pi r \tau \left(\sh\eta{-}\beta_{\ell} \ch\eta \right)y_1^0
   -\int_{x,x'} 4\pi^2 r' \tau \left( \sh\eta{-}\beta_{\ell} \ch\eta \right)
                          \zeta_{01}^0,
\nonumber\\
  R_{\ell s,k}^{2(c)} &=& \int_x 2 \pi r \tau \left(\sh\eta{-}\beta_{\ell} \ch\eta \right)y_1^k
   -\int_{x,x'} 4\pi^2 r' \tau \left( \sh\eta{-}\beta_{\ell} \ch\eta \right)
                          \re\zeta_{01}^k,
\nonumber\\
  R_{\ell s,k}^{2(s)} &=& -\int_x 2 \pi r \tau \left(\sh\eta{-}\beta_{\ell} \ch\eta \right)x_1^k
   -\int_{x,x'} 4\pi^2 r' \tau \left( \sh\eta{-}\beta_{\ell} \ch\eta \right)
                          \im\zeta_{01}^k
\end{eqnarray}
\begin{eqnarray}
\label{eqA7}
  R_{o\ell}^2 &=& \int_x 2 \pi \tau \left(\sh\eta{-}\beta_{\ell} \ch\eta \right)
  \left(r {\mathcal{X}}_1 -\beta_{\perp} \tau\ch\eta {\mathcal{X}}_0 \right)
   -\int_{x,x'} 4\pi^2 \tau \left(\sh\eta{-}\beta_{\ell} \ch\eta \right)
   \left( r' {\mathcal{X}}_0{\mathcal{X}}_1' -\beta_{\perp} \tau' \ch \eta' 
   								{\mathcal{X}}_0{\mathcal{X}}_0'\right),
\\ \label{eqA8}
  R_{o\ell,0}^2 &=& \int_x 2 \pi \tau \left(\sh\eta{-}\beta_{\ell} \ch\eta \right)
  \left(r x_1^0 -\beta_{\perp} \tau\ch\eta x_0^0 \right)
  -\int_{x,x'} 4\pi^2 \tau \left( \sh\eta{-}\beta_{\ell} \ch\eta \right)
  \left( r' \xi_{01}^0 -\beta_{\perp} \tau' \ch \eta' \xi_{00}^0\right),   
\nonumber\\
  R_{o\ell,k}^{2(c)} &=& \int_x 2 \pi \tau\left(\sh\eta{-}\beta_{\ell} \ch\eta \right) 
  \left(r x_1^k -\beta_{\perp} \tau\ch\eta x_0^k \right)
   -\int_{x,x'} 4\pi^2 \tau \left( \sh\eta{-}\beta_{\ell} \ch\eta \right)
   \left( r' \re\xi_{01}^k -\beta_{\perp} \tau' \ch \eta' \re\xi_{00}^k\right),
\nonumber\\
  R_{o\ell,k}^{2(s)} &=& \int_x 2 \pi \tau\left(\sh\eta{-}\beta_{\ell} \ch\eta \right)
  \left(r y_1^k -\beta_{\perp} \tau\ch\eta y_0^k \right)
   -\int_{x,x'} 4\pi^2 \tau\left( \sh\eta{-}\beta_{\ell} \ch\eta \right)
   \left( r' \im\xi_{01}^k -\beta_{\perp} \tau' \ch \eta' \im\xi_{00}^k\right).
\end{eqnarray}
\end{widetext}
%  

%%%%%%%%%%%%%%% References %%%%%%%%%%%%%%%%%%%%%%%%%%%%%%%%%%%%%%%%%%%%%%%%%%%

%\bibliographystyle{utphys}

\end{document}